\documentclass{article}

\usepackage{amsmath}
\usepackage{amssymb}
\usepackage{dsfont}
\usepackage[export]{adjustbox} 
\usepackage{authblk}
\usepackage[pdfborder={0 0 0}]{hyperref} 
\usepackage{float}

\newcommand{\beginsupplement}{%
    \setcounter{table}{0}
    \renewcommand{\thetable}{S\arabic{table}}%
    \setcounter{figure}{0}
    \renewcommand{\thefigure}{S\arabic{figure}}%
} 

\usepackage{longtable}
\usepackage{array}
\usepackage{afterpage}
\newcolumntype{R}[1]{>{\raggedleft\arraybackslash}p{#1}}                                                  
\newcolumntype{L}[1]{>{\raggedright\arraybackslash}p{#1}}

\DeclareMathAlphabet{\mathcal}{OMS}{cmsy}{m}{n}

\title{Aligning Statistical Dynamics Captures Biological Network Functioning}

\author[1]{Ryan E. Langendorf}
\author[2]{Debra S. Goldberg}

\affil[1]{Environmental Studies\\
University of Colorado Boulder\\
\url{ryan.langendorf@colorado.edu}}
\affil[2]{Department of Computer Science\\
University of Colorado Boulder\\
\url{debra@colorado.edu}}

\date{}

\begin{document}

\maketitle


\begin{abstract}
Empirical studies of graphs have contributed enormously to our understanding of complex systems. Known today as network science, what was originally a theoretical study of graphs has grown into a more scientific exploration of communities spanning the physical, biological, and social. However, as the quantity and types of networks have grown so has their heterogeneity in quality and specificity. This has hampered efforts to develop general network theory capable of inferring functioning and predicting dynamics across study systems. We have successfully approached this challenge by aligning networks to each other rather than comparing parameter estimates from individually fitted models or properties of edge topologies. By comparing the predictability of statistical dynamics originating from each network's constituent nodes we were able to build a functional classifier that distinguished underlying processes in both synthetic and real-world network data spanning the entire biological scale from cellular machinery to ecosystems. 

\end{abstract}


\section{Introduction}

Alignment methods attempt to find overlap in pairs of non-Euclidean data as a means to compare complex systems and identify similarities within them. There are many kinds of alignments, but they vary in their adoption and utility across fields and data types. Consider the success of sequence alignments like BLAST \cite{altschul1990basic} which is queried more than 100,000 times every day \cite{mcginnis2004blast}. Network alignments exist, but their adoption has been slow and fractured. This is the result of two prevailing challenges. \textbf{1} Network data varies dramatically in both size and detail, spanning orders of magnitudes in their numbers of nodes and ranging from simple unweighted, undirected networks \cite{watts1998collective} to more realistic ones akin to a system of differential equations \cite{sakamoto2001inferring}. It is precisely this generality of network models that makes network aligners both very tempting to employ and nontrivial to justify. \textbf{2} True global network alignment is NP-complete and quickly becomes computationally intractable as network size increases \cite{clark2014comparison, cook1971complexity} so all aligners rely on heuristics and there is justifiable disagreement on what it means for two networks, or nodes within them, to be similar. Even so, there is simply too much network data on physical systems \cite{aggarwal1997method, kansky1963structure, amaral2000classes}, natural communities \cite{azofeifa2016generative, sharan2006modeling, jeong2001lethality, white1986structure, thebault2010stability}, and social dynamics \cite{kossinets2006empirical, fowler2006legislative, zachary1977information, batty2008size} to not consider the utility of network alignments more carefully. 

The first issue is likely to get worse as network scientists continue to study increasingly different kinds of systems. Then, the key challenge to making network alignments more widely applicable is the difficulty in deciding what it means for networks to be similar. Simple measures of distances such as $\sqrt{\sum_i {(\vec{x}_{1_i}} - \vec{x}_{2_i})^2}$ are unlikely to capture dynamics of coupled nonlinear systems $\vec{x}_{1_i}$ and $\vec{x}_{2_i}$, and often assume the two networks are the same size which is rarely the case. Instead, some additional measure $m(\vec{x}_{1_i},\vec{x}_{2_i})$ is needed to serve as a mapping between the systems. The unresolved question is what $m$ should be. 

This problem was originally couched in graph theory with the goal of creating a bijection between two graphs such that every node in each graph gets paired with exactly one node in the other graph \cite{prager2009historical, bourbaki1966general}. These mapping functions can be used to recover a scrambled graph, but are unable to address degrees of node or graph similarity. More recent developments by network scientists have focused on approximate solutions that emphasize recovering a system's functioning over its exact topology \cite{conte2004thirty}, with some notable successes \cite{milenkovic2008uncovering, nabieva2005whole, kuchaiev2010topological, kuchaiev2011integrative, thebault2010stability}. Even so, judging the quality of an alignment remains a challenge, and most measures emphasize the graph theoretic goal of topological similarity over emergent functional similarity, even when studying the latter. Unsurprisingly then, network alignments have mostly improved at producing topological alignments \cite{clark2014comparison}. It remains less clear how well to expect these approaches to perform at identifying functional similarities or predicting system trajectories in networks modeling complex systems structured by different processes occurring at vastly different spatial and temporal scales.

We have developed a principled and non-topological network aligner to explore the utility of comparing networks by their dynamics rather than properties of their edge topologies. Our approach uses the entropies of simulated diffusion kernels emanating from each node in two networks to make a global pairwise-node alignment between them. We contextualized our method by comparing it to ten recently compared network aligners \cite{clark2014comparison}, explored its behavior relative to known network dynamics, and tested its ability to functionally classify both common synthetic networks as well as biological systems ranging from gene regulatory networks all the way up in scale to food webs. 

\section{Methods}
\subsection{Conceptual rationale}

Details of the algorithm are visually accompanied (see Figure 2) by the following two networks (Figure 1) which emphasize our general approach, of system dynamics mattering more in the actual utility of network alignments than their literal topology which matters more for codifying systems than understanding them. 

\begin{figure}[H]
    \begin{center}
        \includegraphics[scale = 0.35]{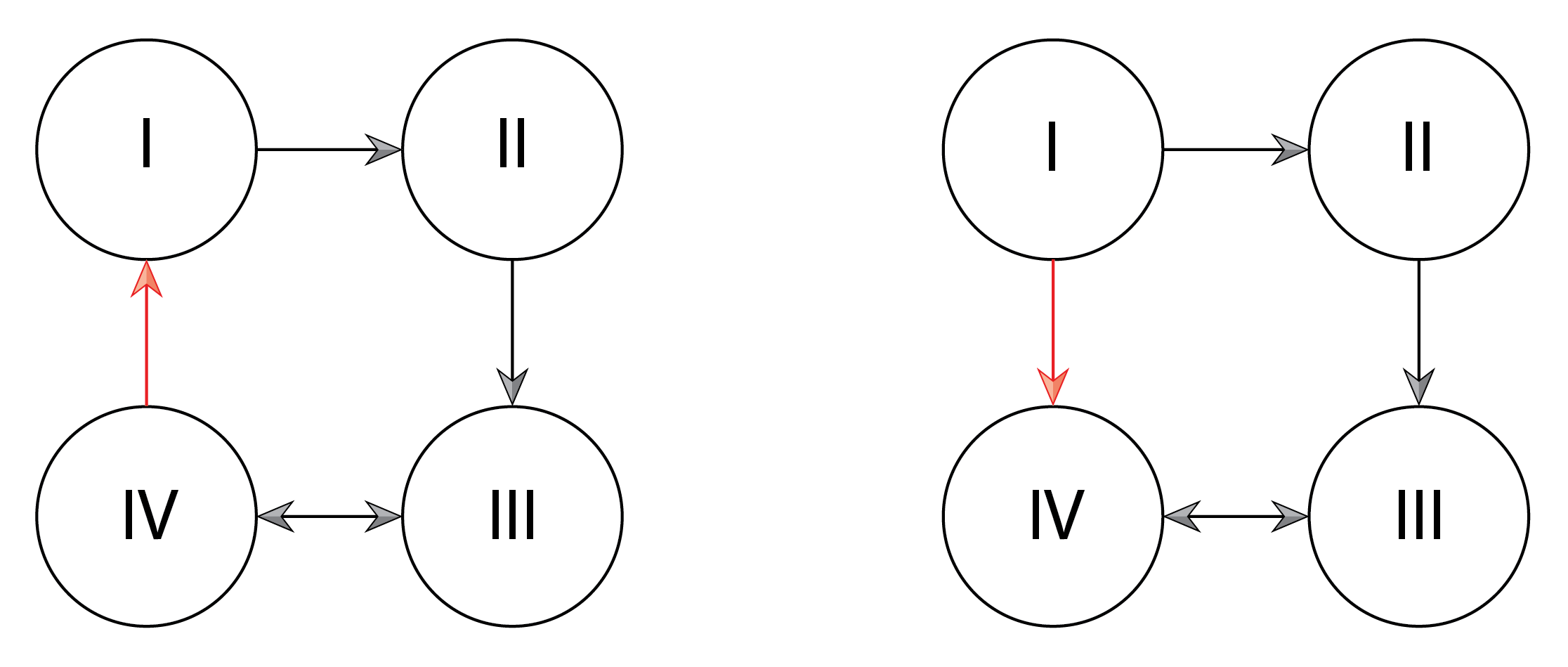}
        \caption{The two networks used as a demonstrative example in the algorithm description. These networks are identical except for the reversal of a single unidirectional edge between nodes I and IV, which is highlighted in red.}
    \end{center}
\end{figure}  

The unidirectional edge between the first and fourth nodes has been reversed, but otherwise these two networks are topologically identical. If aligning two networks is formulated topologically, as a kind of layout problem where the aligner seeks to match up the edges with the minimum number of discrepancies, then these two networks should be classified as nearly identical. We argue that they are in fact quite different, and embody the kind of functioning we hypothesized a dynamics-oriented approach to network alignment would more successfully capture. The left network is a cycle whereas the right network has source-sink dynamics \cite{pulliam1988sources}, which can be seen in their node-specific equilibria where the left network can be in any state but the right network can only be in states III or IV. 

Comparing networks node-wise has this advantage of being able to trace systemic differences to individual components of each system, which is why the approach implemented here relies on capturing the dynamics of the two networks being aligned from the perspective of each node. Consider trying to compare two hypothetical cities of houses connected by roads. Our approach is to pairwise compare each house with those in the other city by creating a house-specific signature. To do so we quantified the predictability of the location of a person at various times after they left their house, assuming they move randomly. This predictability across all houses captures much of the way each city is organized and, we hypothesized, functions. We aligned networks using this conceptual rationale, with nodes as houses, edges as roads, and random diffusion representing people leaving their houses and walking around the city to other houses. The mechanics of this, which are conceptually akin to flow algorithms \cite{nabieva2005whole} and Laplacian dynamics \cite{mucha2010community}, are depicted in an example alignment in Figure 2 using the two networks in Figure 1.

\begin{figure}[H]
\begin{center} 
    \includegraphics[max width=\linewidth]{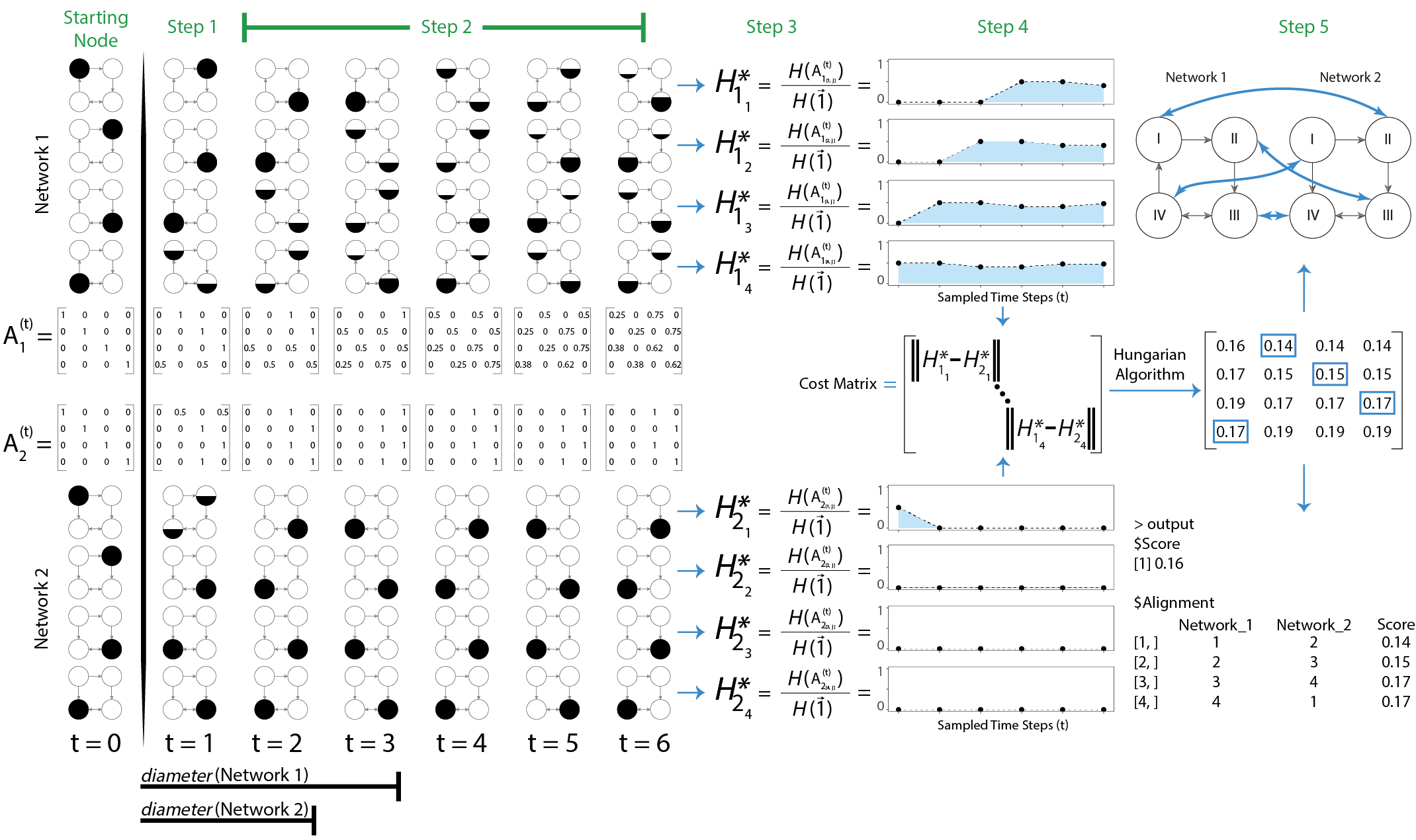} 
    \caption{Aligning two networks by comparing the predictability of dynamics originating from their constituent nodes. Analytically this is equivalent to (Step 1) turning a network into a Markov process by row-normalizing its matrix representation to unity, (Step 2) exponentiating this Markov process matrix to different powers, which is identical to sampling the diffusion kernel at different time steps, (Step 3) quantifying the predictability of the system across these time steps using a normalized version of Shannon's entropy, (Step 4) storing in a cost matrix the numerically integrated difference between these predictability-over-time curves for all pairwise starting points in the two networks, and finally (Step 5) using the Hungarian algorithm to find an optimal alignment between nodes based on how the entropy of the system changed over time for diffusion emanating from each node. Sample output from the R package netcom is presented at the bottom right.} 
\end{center}
\end{figure}

\subsection{Networks}

A network (or graph) $G$ was defined as $N$ nodes connected by $E$ edges of potentially varying weights determined by a mapping function $w:E \rightarrow \mathbb{R}$. That is, $G:=(N, E_w)$. Note that we did not consider node properties. These networks can also be represented as a transition matrix $A$, where the element $A_{ij}$ is the edge from node $i$ to node $j$ and an edge of weight zero indicates no edge.  

We used only static networks. While dynamic edge and node functions require a more simulation-oriented approach than that developed here, the general method is amenable to additional input functions that specify how edges and/or nodes change as a function of time and/or each other, which other network aligners are not. Similarly, this study used only connected networks, but the method can also be applied to networks comprised of disconnected subgraphs between which all edges have weight zero. 

\subsection{Alignment algorithm}

This family of network alignment algorithms proceeds in 5 general steps to produce a node-level pairwise alignment between two input networks. This creates a bijection between the nodes of two identically sized networks and an injection if one is larger. No single model can perfectly characterize a network, which is why this method should be viewed as a family of flexible methods that all align two networks using the same conceptual rationale, but which can differ in their implementations to accommodate varying network types as well as analytic goals. We have distinguished these \textit{Options} from the core algorithm and indicated our implementation. 

\subsubsection{Step 1/5} Convert the $N_1$x$N_1$ and $N_2$x$N_2$ square transition matrix representations of the two networks being compared into Markov processes by normalizing each row sum to unity \cite{durrett2012essentials}. For each network's new transition matrix ${A_x}^{(1)}$, $\sum_{j=1}^{N_x} {A_{x_{ij}}}^{(1)} = 1$ $\forall$ $i \in {1,2,...,N_x}$. The $i,j_{\text{th}}$ position is then the probability of moving from the $i_{th}$ component of the network to the $j_{th}$ in a single unit of time, from here on denoted by parenthetical powers as ${A_x}^{(1)}$. 

Empirical networks vary greatly in their size, reflecting a range of system complexity but also an inherent noise in the current network data. We chose to allow differently sized networks to be compared, but with an optional penalty created by adding disconnected nodes to the smaller network colloquially called padding. This approach allows the use of common assignment methods like the Hungarian algorithm which require a square cost matrix (2.3.5 Step 5/5), and creates a penalty that grows approximately linearly with the gap in size as these additional nodes will always align poorly to nodes of degree $>$ 0. Moreover, this penalty can be ignored without changing the alignment scores between actual nodes in the networks, as was done in all of our analyses, to make size-invariant comparisons of system functioning. 

\bigskip \noindent \textit{Option:} Normalizing each row to unity will alter the ratio of edge weights between nodes with different out-degrees. This can be prevented by adding a row-specific constant to each diagonal element ${A_{x_{ii}}}^{(1)}$ such that $\sum_j {A_{x_{ij}}}^{(1)} = K$ for all nodes $i$ in the network. We did not use this normalization because our analyses included only unweighted networks where edges are nominally present or absent and are not considered comparable in strength.

\bigskip \noindent \textit{Option:} Note that for directed networks we use the row-to-column orientation of the adjacency matrix in the style of Ulanowicz \cite{kay1989detailed}, but column-to-row matrix models can be employed simply by transposing the adjacency matrices ${A_x}^{(1)}$ before proceeding. Our implementation creates a signature for each node by mapping diffusion emanating outward, but depending on the application it might make more sense to capture how diffusion kernels arrive at a given node. A linear combination of both can be considered, though this requires exponentiating both ${A_x}^{(1)}$ and ${{A_x}^T}^{(1)}$ adding to the computational complexity of the algorithm. For smaller analyses where the investigator is unsure which interpretation better applies, $\alpha {A_x}^{(t)} + \beta {{A_x}^T}^{(t)}$ should provide a more generally unbiased alignment, where $\alpha + \beta = 1$ and $\alpha$, $\beta > 0$. 

\subsubsection{Step 2/5} Separately exponentiate the adjacency matrices to different powers $t \in \mathbb{N}$, where the resultant matrices ${A_x}^{(t)}$ are the probabilities of interactions transferring some currency (e.g. energy, information) from the $i_{th}$ member of the system to the $j_{th}$ in exactly $t$ units of time. In this way enumerating different powers of $A$ is equivalent to discretely sampling a diffusion kernel. 

\bigskip \noindent \textit{Option:} Which powers $A$ should be raised to is application specific. We have used $t \in \{2^y\}$ $\forall$ $y \in \mathbb{N}$ such that $t_{max} \geq 2D$ and $t_{max-1} < 2D$ where $D$ is the larger diameter of the two networks being compared. That is, the largest element of $t$ was always at least twice the network's diameter. This set of time steps was chosen to ensure every cycle could be traversed at least once. Additionally, sampling logarithmically helped preserve the uniqueness of the origin node as many diffusion patterns tend toward the same stationary distribution regardless of their initial configuration, while emphasizing direct and local interactions that may matter more in real-world functioning. This is similar to sampling every time step with a non-uniform weighting function, but is computationally less expensive. Note that this method of exponentiating the normalized transition matrix only works for $t \in \mathbb{N}$. Continuous-time Markov models \cite{spencer2005continuous} can be used to model fractional matrix powers, but the added complexity is less worthwhile here where each system's state is sampled discretely regardless.  

\subsubsection{Step 3/5} The diffusion kernels generated by Step 2 need to be compared. For the two example networks in Figure 2, their kernels could be directly compared in Euclidean space. However this limits the method to aligning networks of the same size. Instead, we used the normalized entropy \cite{shannon2001mathematical} of each row of each ${A_x}^{(t)}$ matrix, creating a sequence of $T$ entropies for each node in each network describing diffusion emanating from that particular node. All together, this produces a characteristic $N_x$x$T$ matrix for each network, here denoted $S_x$, where each node's signature entropy-over-time curve is the same length allowing nodes in differently sized networks to be compared.

\begin{equation} \label{eq1}
   S_{x_{n,t}} = \frac{-\sum_{j=1}^{N_x} A_{x_{n,j}}^{(t)}\text{ln}(A_{x_{n,j}}^{(t)})}{H(\mathds{1}^{N_x})} 
\end{equation}

\noindent Note that the entropies were size-normalized and only calculated including nodes originally in each network, even for those diffusion kernels emanating from padding nodes added to the smaller network. The size normalization $H(\mathds{1}^{N_x})$ is important to prevent incorporating the size bias inherent in Shannon's formulation of entropy. To see this bias, consider the entropies of $\{1,1\}$ and $\{1,1,1\}$. Entropy increases with size because the added states the system can be in reduces information content and predictability. We therefore normalized all entropies by the maximum entropy of a system of the same size. This normalizing constant can be thought of as the null model, where the system is entirely random and maximally unpredictable. 

\bigskip \noindent \textit{Option:} The diffusion kernels can be captured using a measure other than entropy. Here we treat network functioning as statistical non-randomness, quantified by random diffusion within the network. This has the advantage of both a rigorous theoretical foundation and demonstrated applicability across multiple orders of scale \cite{fergus2003object, avci2007expert, jurie2004scale}. Other measures (e.g. 1-D: Simpson index \cite{simpson1949measurement}, Gini coefficient \cite{gini1912variabilita}; $>$1-D: ordination techniques) can and should be adopted depending on the systems being compared and the intended use of the alignment. 

\subsubsection{Step 4/5} Store the Euclidean distance between rows of the two matrices $S_1$ and $S_2$ in a square cost (or distance) matrix $C$ with dimensions equal to the number of nodes in the larger of the two networks being aligned. The diagonals of $C$ will always be zero, and it will always be off-diagonal symmetric.  

\begin{equation} \label{eq2}
\begin{split}
    C_{ab} &= \sqrt{ \sum_{t \in T} {(S_{x_{1_{a,t}}} - S_{x_{2_{b,t}}})}^2 }
\end{split}
\end{equation}

\subsubsection{Step 5/5} Run the Hungarian algorithm \cite{konig1916graphen, egervary1931matrixok, kuhn1955hungarian, munkres1957algorithms, kuhn2012tale} on the cost matrix $C$ to find the optimal node-level alignment between the two networks. The arithmetic mean of these pairwise node alignments was considered the overall alignment score between the two networks. 

\bigskip \noindent \textit{Option:} The size penalty for aligning networks with differing numbers of nodes can be assessed here by including the full bijection output from the Hungarian algorithm. We chose to ignore this penalty in our analyses to reduce any size bias associated with heterogeneous network data, and therefore only averaged the elements of the cost matrix associated with nodes originally in both networks.

\bigskip \noindent \textit{Option:} The Hungarian algorithm guarantees an optimal solution to the assignment problem, and therefore the best possible alignment given the cost matrix $C$ from Step 4. However, it's $\mathcal{O}(n^3)$ runtime makes it computationally expensive for applications involving very large networks. To limit biases in our analyses, and because matrix multiplication is almost as expensive a process \cite{davie2013improved}, we exclusively used the Hungarian algorithm. The cost matrix $C$ can however be fed into any of the many methods of solving the assignment problem that use heuristics to run faster \cite{clark2014comparison}. 

\bigskip \noindent \textit{Option:} It will often be more appropriate to consider the probability of having made an alignment of the same quality rather than the raw alignment score. Consider obtaining an alignment score of 0.01. It is unclear if this is a surprisingly good alignment or inevitable given the types of networks under consideration. Testing for this requires creating stochastic versions of the two networks under consideration which can then be aligned to bootstrap an alignment distribution from which a p-value can be empirically calculated. Rather than introduce the bias of choosing a means of stochastically generating similar networks, and to avoid the added computational complexity, we used raw alignment scores in our analyses. However there is a need for null models in the alignment literature that deserves more attention if network aligners are to be applied more commonly across studies and even fields.

\subsection{Analyses}

Alignments produced by this method need to be reliable such that surprises can be thought of as hypothesis generating rather than methodological artifacts. We attempted to asses this reliability rigorously using synthetic data in several ways.

\subsubsection{Comparison with other aligners}
Given the computational complexity of matrix exponentiation, if our approach is not meaningfully different from current network aligners why not continue developing the same kinds of functional inference with an already established method? Clark and Kalita \cite{clark2014comparison} recently used a subset of the synthetic protein-protein interaction (PPI) networks in the NAPAbench database, which was created to test network alignments \cite{sahraeian2012network}, to compare ten common network aligners. As this subset of the network alignment literature is perhaps the most robust and functionally-oriented, we chose to focus on the same methods in providing context. We calculated the same Induced Conserved Structure (ICS) \cite{patro2012global} of the alignments produced by the approach detailed here and compared them to the corresponding scores Clark and Kalita \cite{clark2014comparison} presented. 

\begin{equation} \label{eq3}
    \text{ICS} = \frac{|f(E_1) \cap f(E_2)|}{|E_{G_2[f(N_1)]}|}
\end{equation}

The numerator of ICS is the number of edges preserved under the alignment, and the denominator is the number of edges in the subgraph of the second network $G_2$ induced by the mapping from the first network $G_1$ (those edges between nodes in $G_2$ mapped to by nodes in $G_1$). This not only measures how many edges are preserved by an alignment, but penalizes aligning sparse and dense regions which can lead to better alignments through combinatorics alone. There were 30 alignments total, 10 each within the following three generative network models: duplication with random mutation \cite{pastor2003evolving}, duplication-mutation-complementation \cite{vazquez2002modeling}, and crystal growth \cite{kim2008age}. In all 30 cases the alignment was made between a network with 3000 nodes and a network with 4000 nodes.

\subsubsection{Node centralities}
One of the most common characterizations of a network is the distribution of the centralities of its nodes and/or edges \cite{newman2010networks}. These characterizations help identify systems that are driven primarily by a few very important components, compared to more neutral systems \cite{hubbell2001unified} where nodes are mostly similar. While alignment algorithms are not primarily measures of centrality, they can be used to measure the importance of a node or edge by aligning the original network to itself after removing one, similar to knocking out a gene \cite{evans2001cultural, hall2009overview} or locally excluding a species \cite{paine1966food}. We compared this alignment-based centrality with 3 common ones (degree, eigenvector, and betweenness), Katz for its conceptual similarity \cite{newman2010networks, katz1953new}, and PageRank for its ubiquitous familiarity \cite{page1999pagerank}. Each of the 200 nodes considered were uniformly randomly selected from randomly generated directed Erdos-Renyi networks \cite{erdos1959random} with 1000 nodes. Half of these were for a noisy static edge density of $p \sim \mathcal{N}(0.3, 0.01)$, whereas the other half had a uniformly random edge density of $p \sim \mathcal{U}(0.2, 0.8)$. This allowed us to explore the relative similarity of node importance between different centralities across a realistic range of model complexity.

\subsubsection{Tracking network dynamics}
More similar networks should align better, with alignment scores closer to zero. Here we defined similarity as an edge-based Hamming distance \cite{hamming1950error}, where the number of edges required to turn one network into the other was used as a baseline measure of network similarity. This allowed us to test the alignment's behavior under known, albeit synthetic, conditions. We simulated 100 directed Erdos-Renyi networks with 100 nodes and an edge density of 0.5, and then randomly removed or added edges before aligning back to the original network. This was done 100 times (once for each network) up to 128 edges at $log_2$ intervals to allow for more replication without sacrificing too much resolution at the likely more biologically relevant smaller levels of change. Additionally, to test the robustness of alignments to network dynamics where the order of lost or gained interactions can matter \cite{drake1991community, collinge2009transient}, edges were removed or added in 3 ways using their respective edge betweenness as a measure of importance \cite{girvan2002community}: randomly, from least to greatest importance, and from greatest to least importance. The importance of each edge was recalculated following each removal or addition to account for systemic changes created by removing or adding edges. 

\bigskip \noindent We also created two network trajectories to demonstrate how an evolving network can be tracked by aligning all pairs of time steps. The first was a 100 node Erdos-Renyi random network \cite{erdos1959random} with edge probability 0.5 that at each time step randomly flipped a single edge from present to absent or absent to present. The second was a 100 node Barabasi-Albert Preferential Attachment random network \cite{albert2002statistical} which was randomly scrambled into an Erdos-Renyi network. At each time step an edge was uniformly randomly picked and then uniformly randomly labeled as present or absent regardless of its previous state. Note that the same edge could be picked multiple times.    

\subsubsection{Functional network classification}
We explored the generality of our alignment algorithm's ability to distinguish different kinds of systems functionally by pairwise aligning 307 empirical networks spanning micro and macro-biology with 120 reference networks generated from four common theoretical models (30 each across a range of sizes). All 427 networks are listed in Table S3. Note that networks were first divided into connected components, if necessary, before being aligned.

For network alignments to be a useful kind of inference they should be able to identify functionally similar communities that are likely to react similarly to the same stimulus. That is, they ought to be able to perform supervised learning to classify unknown networks functionally. To test this we used synthetic data to classify unknown networks solely by aligning them to known networks. All networks were generated from one of four commonly-cited generative models: directed Erdos-Renyi \cite{erdos1959random}, Barabasi-Albert Preferential Attachment \cite{albert2002statistical}, Duplication and Divergence \cite{ispolatov2005duplication}, and Watts-Strogatz Small World \cite{watts1998collective} networks. Using the Variable Version scheme listed in Table S1 to allow for conservatively noisy data, we generated 1000 networks with exactly 100 nodes, uniformly randomly according to one of the four models. This process was then repeated. Each of the 1000 new networks was aligned to all 1000 of the previously generated known networks. First the model type was predicted as the minimum of the average distance to all networks within each type. Then, the parameter itself was predicted as an average of the parameter values of those networks belonging to the predicted type weighted exponentially by their alignment scores with the unknown network. An exponential weight of $\lambda = 100$ was used to place most of the predictive power on those networks that aligned well with the unknown network because network similarity is likely nonlinear with respect to underlying parameters. As most empirical networks vary in size even when describing the same process, we repeated this analysis with networks having an integer number of nodes ranging uniformly randomly between 10 and 100. The actual and predicted model type and parameter were then plotted against each other (Figure 7). 

To explore the real-world utility of aligning networks by simulating their dynamics we attempted to classify 100 ecological networks assembled by the Systems Ecology and Ecoinformatics laboratory at the University of North Carolina Wilmington \cite{borrett2014enar}. This subset of the 307 empirical networks in Figure 6 were annotated both traditionally, by their composition, and functionally allowing us to test our approach's ability to recover functional classifications. Additionally, they were assembled by the same research group limiting differences in the nature of the models. We chose to focus on ecological communities both for their age, as ecologists have been using networks to describe communities for decades \cite{deangelis1983current, strong1988special, hastings2016introduction}, and for their inherent noise \cite{novak2011predicting} as a way to ensure conservative results. In addition to the pairwise alignments, we also attempted to ascertain the underlying processes in each network by aligning them to randomly generated networks of the same size from each of the four reference network models as in Figure 7. These networks ranged from 4 to 125 nodes with an average and standard deviation of 26.66 $\pm$ 27.14. Each model's parameter state space was divided evenly into 101 bins (including the endpoints), and each of these models was used to randomly generate 100 networks. A dimension reduced plot of the pairwise alignment scores between these 100 ecological networks along with the averages of the alignments to the four reference models are plotted in Figure 8. A Shepard plot of the deviations in the dimension reduced plot is given in Figure S4.

\subsection{Data availability}
All biological networks used in this study were publicly available as of February 20, 2017 and listed in Table S3. All reference networks were simulated using the R igraph package \cite{iGraph}, version 1.0.1, except for the Duplication and Divergence networks which were generated as in \cite{ispolatov2005duplication} but with an additional parameter governing the probability of a duplicated node linking to the original node \cite{vazquez2002modeling, gibson2011improving}. See Table S1 for details. 

\subsection{Code availability} The alignment algorithm developed here is part of the R package netcom \cite{NetCom, R}. All analyses were run using the following alignment parameters: netcom::align(..., ..., base = 2, characterization = ``entropy'', normalization = FALSE)

\section{Results}

\subsection{Comparison with other aligners} 
Alignments produced by our method were fundamentally different than any of those produced by the ten algorithms Clark and Kalita \cite{clark2014comparison} studied (all alignment scores are listed in Table S2). The highest individual alignment's ICS score across all types was only 0.046, in contrast to all ten network alignments Clark and Kalita \cite{clark2014comparison} tested which mostly produced scores above 0.5. Natalie 2.0 \cite{el2011lagrangian} even managed to average above 0.8. No other aligner, even the older GRAAL \cite{kuchaiev2010topological} which performed the worst, produced ICS scores below 0.1. The differences in ICS scores produced by the method developed here suggest current aligners preserve network topology at the expense of similarities in their dynamics.

\subsection{Node centralities}
Despite producing novel alignments, the alignment scores of Erdos-Renyi networks aligned with themselves following the removal of a single node were positively correlated with common measures of centrality (degree, eigenvector, betweenness, Katz \cite{newman2010networks, katz1953new}, and PageRank \cite{page1999pagerank}) when the networks had a noisy but static edge probability (Figure 3). However, under a uniformly random model these correlations weakened and even became negative (betweenness). This disagreement was captured by the empirical centrality distributions of each measure, which are presented along the diagonal in Figure 3. When the edge probability was noisily static all six centralities were unimodal and qualitatively symmetric. However when that parameter was allowed to vary randomly they behaved differently. Degree centralities became more uniform, eigenvector and betweenness centralities skewed toward more and less important nodes respectively, Katz centralities shifted toward less important nodes but remained symmetric, and PageRank and the alignment-based centrality remained symmetric and unimodal about approximately the same mean node importance. This centrality invariance with respect to the variability of underlying parameters makes the approach developed here more amenable to between-study comparisons and extrapolations of node importance that deserves further study.  

\begin{figure}[H]
    \begin{center}
        \includegraphics[max width=\linewidth]{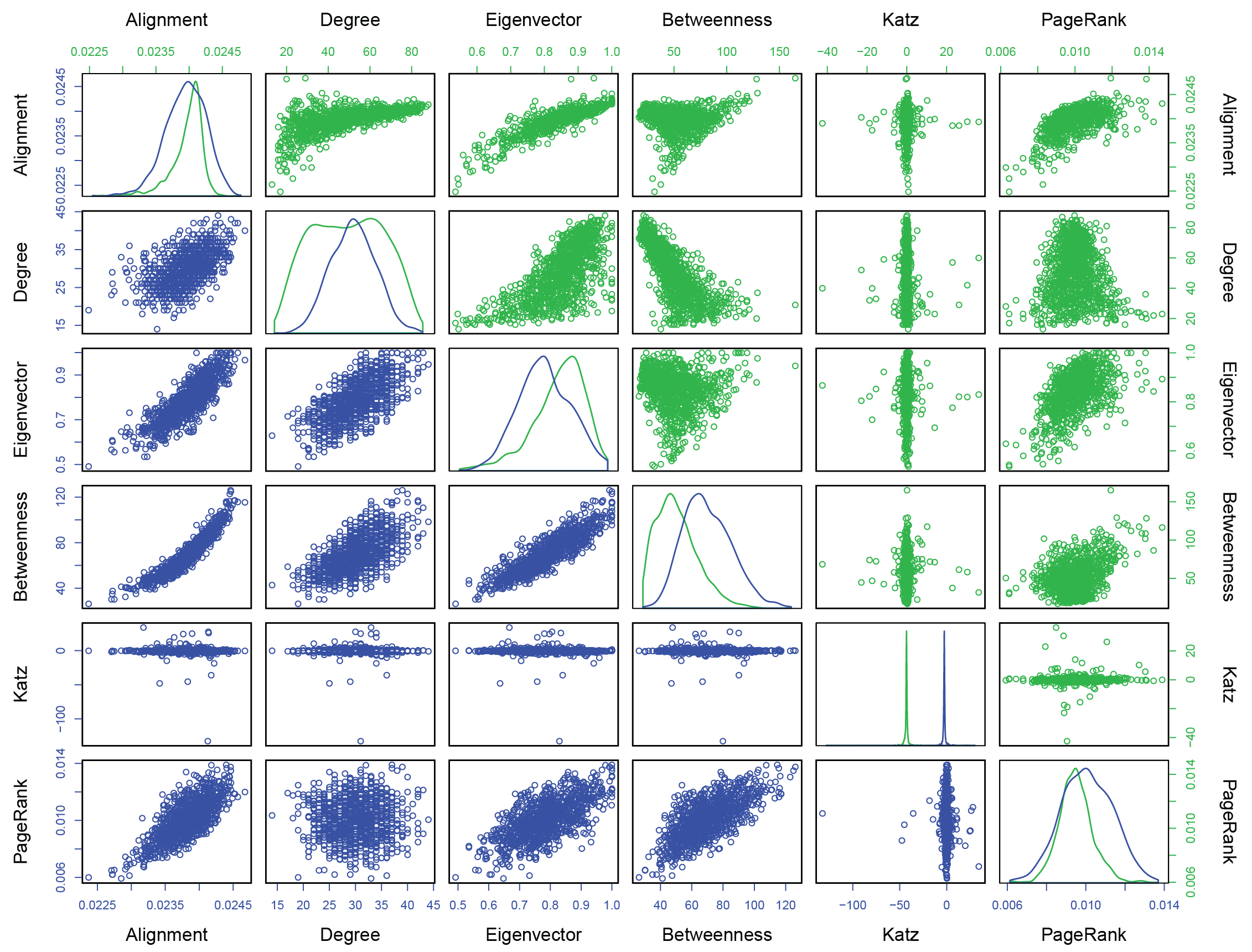} 
        \caption{A comparison of five common measures of node importance with each other and the diffusion-based alignment algorithm presented here (Alignment). Each of the 1000 points per panel was a randomly selected node from a randomly generated 100 node directed Erdos-Renyi network. The lower diagonal (blue) networks had a stochastic edge probability of $p \sim \mathcal{N}(0.3,0.01)$ and the upper diagonal (green) networks had a stochastic edge probability of $p \sim \mathcal{U}(0.2,0.8)$. The diagonal contains the numerically smoothed empirical centrality distributions, with the same blue/green colorings. Note the color-coded centrality-specific axes.}
    \end{center}
\end{figure}

\subsection{Tracking network dynamics}
Alignments monotonically worsened, though not symmetrically, as more edges were removed or added independent of whether this was done randomly or in order of edge betweenness (Figure 4). This held across three orders of magnitude in the number of edges changed suggesting the potential to track systemic changes by aligning a network to itself through time. This approach may help identify critical points as well as assess general system stability, as demonstrated in Figure 5, and can be equivalently applied along other dimensions (e.g. spatial) if the underlying dynamic is not temporal in nature. See Figure S1 for a Shepard plot of all deviations from the true alignment scores between each pair of time steps.

\begin{figure}[H]
    \begin{center}
        \includegraphics[max width=\linewidth]{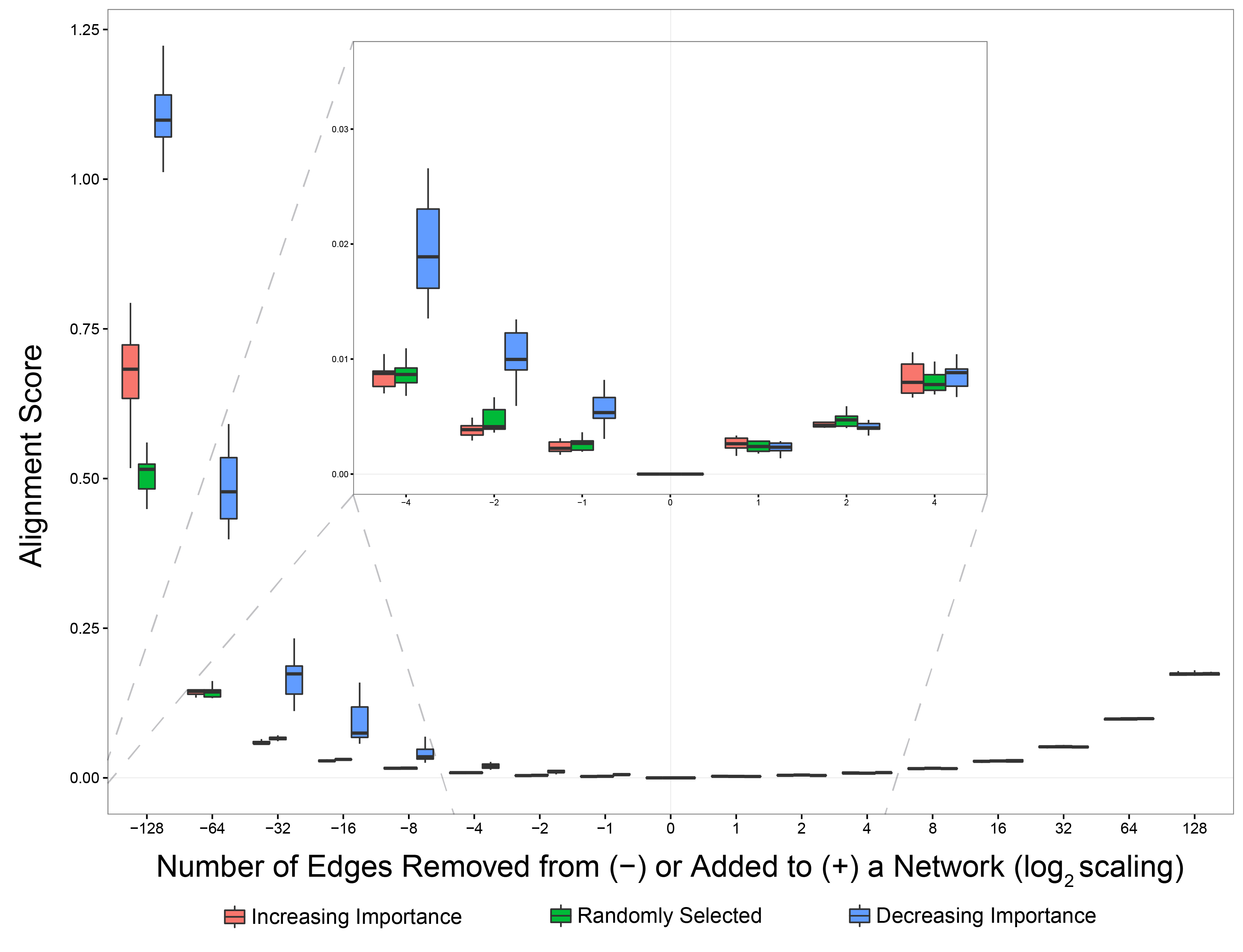} 
        \caption{The change in the alignment score of a network aligned with itself following edge removals or additions. All networks began as 100 node directed Erdos-Renyi networks with an edge density of 0.5. Edges were then removed (negative x-axis values) or added (positive x-axis values) randomly (green), starting with the most important (red), or least important (blue). Boxes show the median, interquartile range (IQR), and confidence interval (1.5*IQR/sqrt(n)). n = 100 networks per box.}
    \end{center}
\end{figure}

\begin{figure}[H]
    \begin{center}
        \includegraphics[max width=\linewidth]{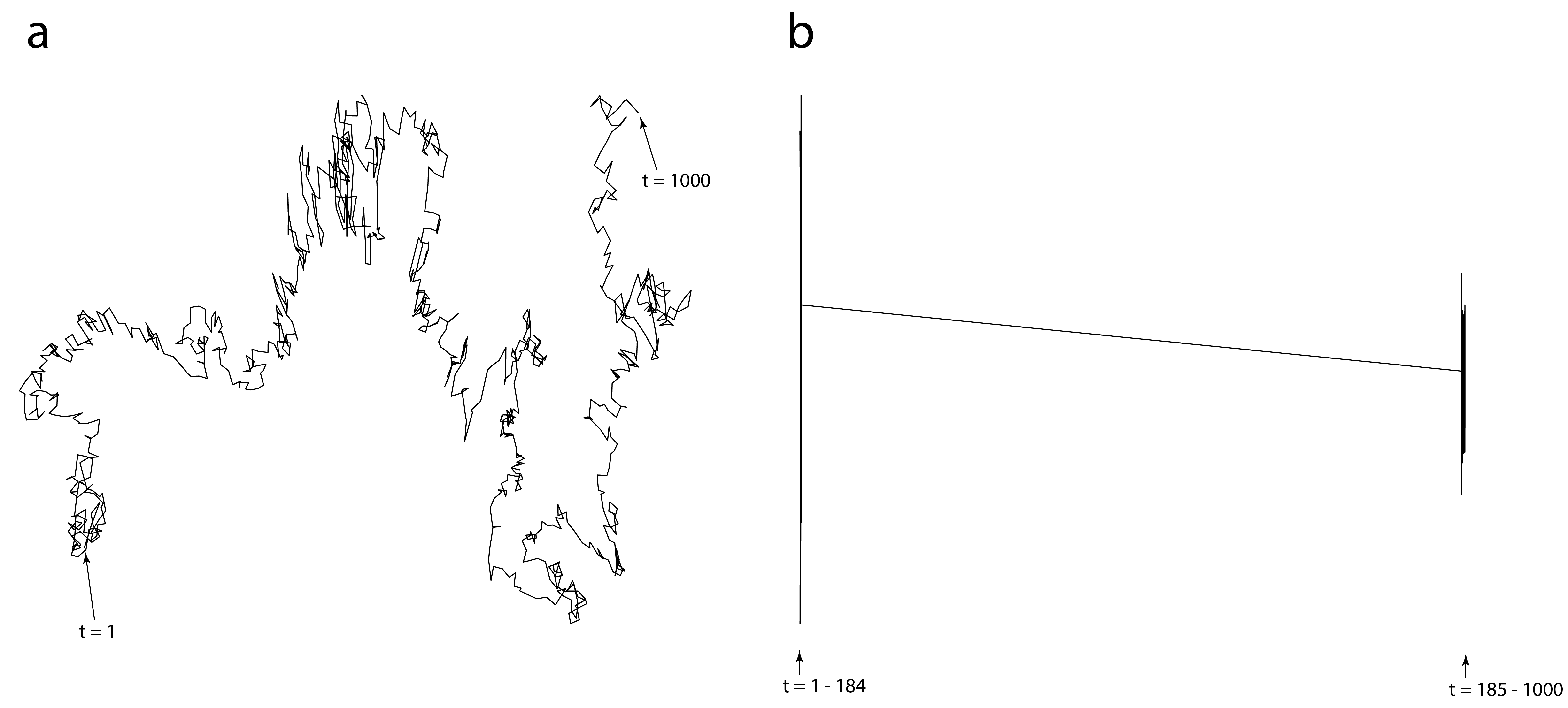} 
        \caption{Non-metric multidimensional scaling of network dynamics captured by aligning all pairs of time steps. \textbf{a)} An Erdos-Renyi network (N = 100, p = 0.5) randomly changing a single edge at each of the 1000 time steps. Stress = 0.07. \textbf{b)} A 100 node network formed by linear preferential attachment randomly changing a single edge at each of the 1000 time steps. Stress = 0.0002. See Figure S1 for Shepard plots of all deviations.}
    \end{center}
\end{figure}

\subsection{Functional network classification}
Comparing networks using their simulated dynamics created novel alignments, correlated with well-studied measures of node importance, and behaved smoothly with respect to underlying changes in a network. We therefore hypothesized our approach would be capable of identifying functionally similar communities that are likely to react similarly to a given stimulus. As seen in Figure 6 our approach was largely able to distinguish between 13 types of systems present in an assembled database of 427 networks spanning micro- and macro-biology (see Table S3 for details). Additionally, the difference in the number of nodes between two networks was unrelated to their alignment score (Figure S3) indicating that our prioritization of dynamics over topology produced size-invariant alignments. Figure S2 contains the corresponding Shepard plot of all deviations in Figure 6 from the true higher-dimension alignment scores.  

\begin{figure}[H]
    \begin{center}
        \includegraphics[max width=\linewidth]{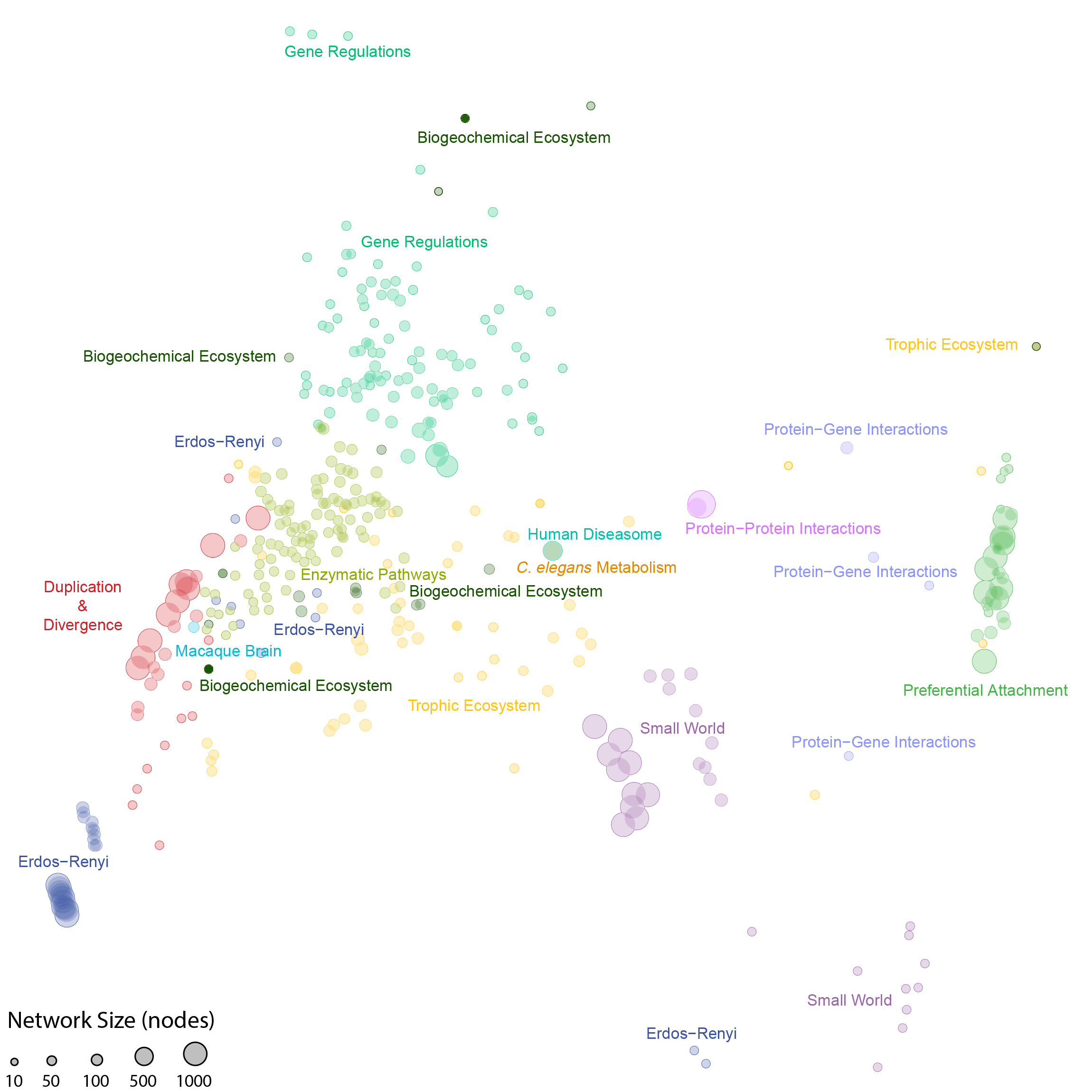} 
        \caption{Non-metric multidimensional scaling of pairwise alignment scores between 307 biological networks and 120 common generative network models, totaling 13 distinct types of systems. Each point is a single network with its size proportional to its number of nodes. The four synthetic types are: Erdos-Renyi, Preferential Attachment, Duplication and Divergence, and Small World. The nine biological types are: enzymatic pathways, gene regulations, protein-gene interactions, protein-protein interactions, C. elegans metabolism, a Macaque's brain (visuotactile regions), the human diseasome, trophic ecosystems, and biogeochemical ecosystems. Stress = 0.11 (see Figure S2 for a Shepard plot of all deviations).}
    \end{center}
\end{figure}

We leveraged our approach's ability to differentiate network types in a classifier that was largely successful at inferring the model and parameter of unknown networks by aligning them to known networks (Figure 7). When network size was fixed every network type was correctly predicted, as seen by all points lying within the diagonal squares. Even when network size was allowed to vary, only 18 networks (1.8\%) were assigned to the wrong model type. All of these networks except for one had fewer than 30 nodes and were Erdos-Renyi models, indicating that smaller random networks may appear as other types of developing systems in their functioning. In general the alignments were most successful at within-model parameter identification of Erdos-Renyi networks, and struggled to identify preferential attachment and small world networks with high probabilities of attachment and rewiring respectively.

\begin{figure}[H]
    \begin{center}
        \includegraphics[max width=\linewidth]{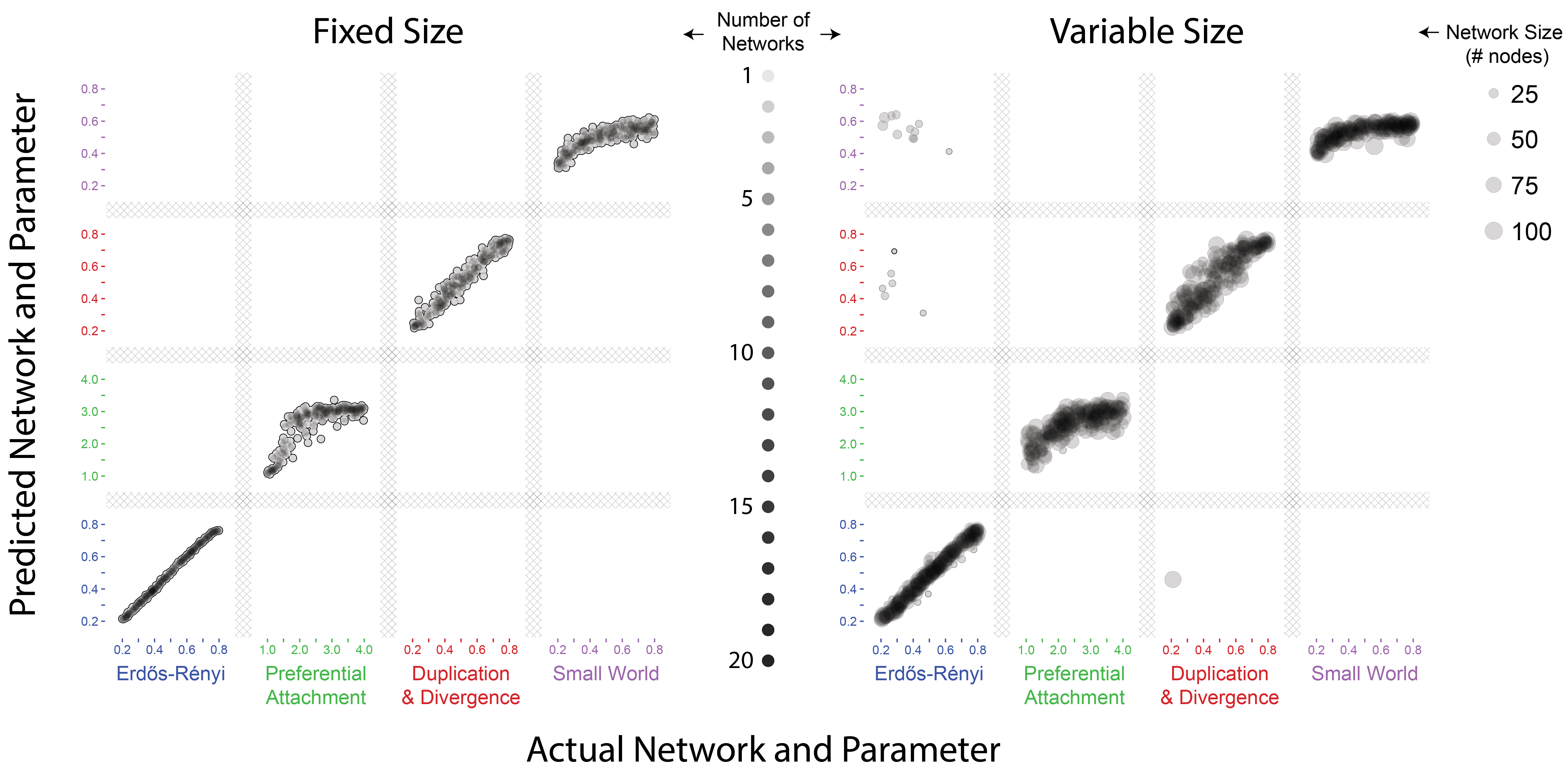}  
        \caption{Classification and parameter prediction of unknown networks. All of the 1000 networks in each panel were created from one of four generative models, with a single uniformly random parameter: Erdos-Renyi's edge density, Preferential Attachment's attachment power, Duplication and Divergence's probability of divergence, and Small World's probability of rewiring. Each network was classified by aligning it with 1000 known networks and averaging their parameters weighted exponentially by their alignment score ($\lambda = 100$). All 1000 networks in the left panel and the 1000 networks they were each aligned to had exactly 100 nodes, whereas in the right panel these networks had a uniformly random integer number of nodes between 10 and 100. There are no networks in the diamond shading which serves only to demarcate the four network types.}
    \end{center}
\end{figure}

We applied this classifier to a subset of the biological networks that were curated by the same research group \cite{borrett2014enar} and labeled with traditional as well as functional classifications which we successfully recovered (Figure 8). This analysis assumes these 100 ecological networks derive from some combination of only four models, which is unknowable in this context and almost surely incorrect. However, these alignments are able to infer if any of the proposed models account for their dynamics. The magnitudes of the alignment scores are of less use here because they are not reliably normalized, but changes in alignment scores following changes in a model's parameter offer evidence of that process being at play in the network being aligned. As an example, the cycling estuary networks all aligned better with Erdos-Renyi networks as the probability of an edge increased, and aligned worse with Duplication and Divergence networks as the probability of divergence increased. In contrast to this they aligned consistently well and poorly with Small World and Preferential Attachment networks respectively. This consistency across parameter values indicates that particular parameter is unrelated to the functioning of the estuary communities, and that any high quality alignments occurred by chance or a shared similarity to a different community driver. In this way, much like regressing covariates, generative models can be assessed as predictors of network data.

\begin{figure}[H]
    \begin{center}
        \includegraphics[max width=\linewidth]{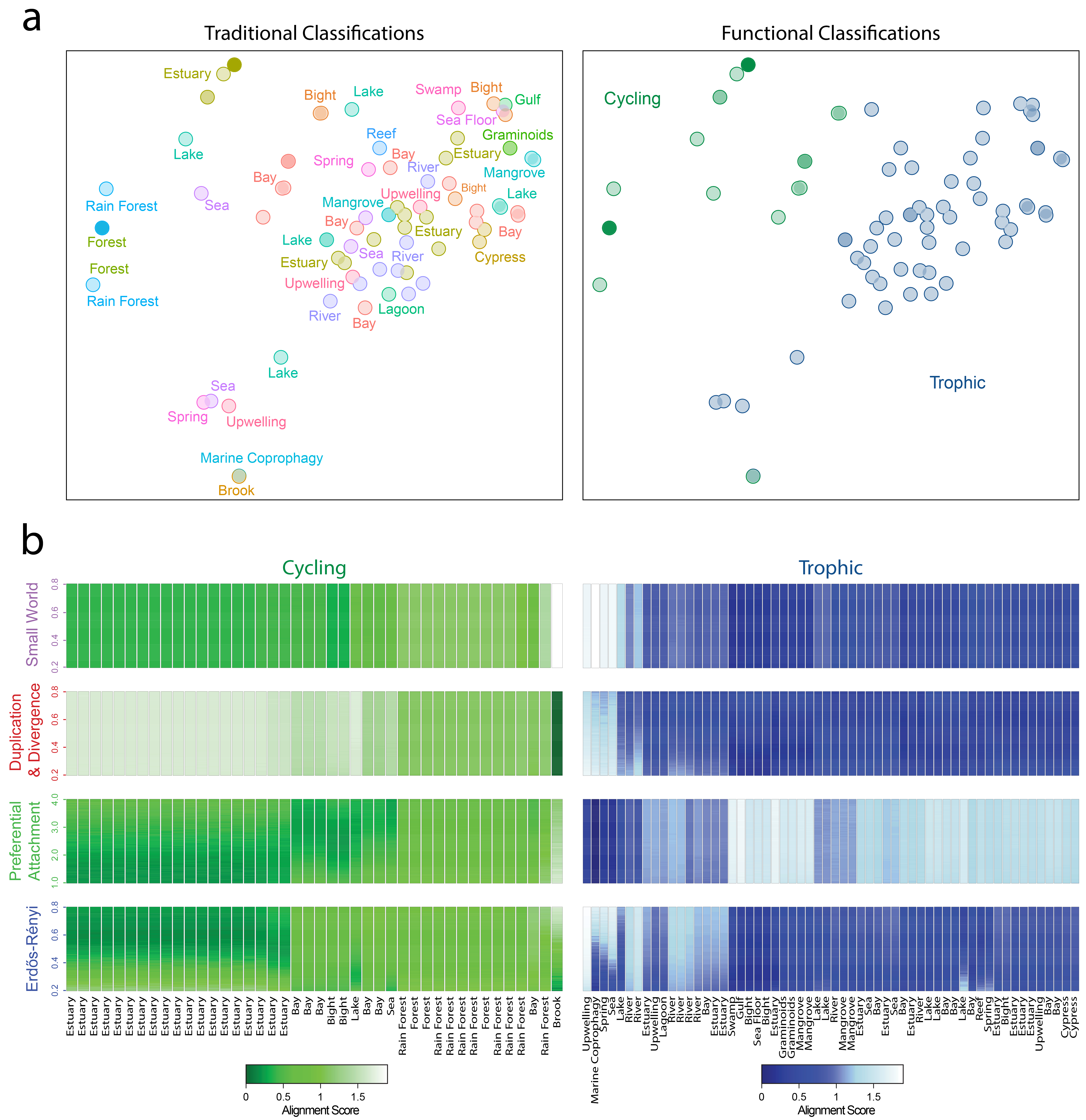}
        \caption{Classification and model prediction of the 100 ecological networks in the enaR database \cite{borrett2014enar}. \textbf{a)} Non-metric multidimensional scaling of the pairwise alignment scores between all 100 networks. These plots are identical except for their labeling and coloring, which indicate the traditional classifications (left) and functional classifications (right) as labeled by their curators at the Systems Ecology and Ecoinformatics laboratory at the University of North Carolina Wilmington. Stress = 0.16 (see Figure S4 for a Shepard plot of all deviations). \textbf{b)} Average alignment scores of each network (column) with randomly generated networks from each of the four theoretical reference models in Figure 7 (see Table S1 for details). The parameter space for each of the network models was divided evenly into 101 bins (rows), each of which was used to create 100 reference networks which were aligned to the ecological networks. The parameters were: Erdos-Renyi's edge density, Preferential Attachment's attachment power, Duplication and Divergence's probability of divergence, and Small World's probability of rewiring. Networks are arranged within the two classifications to keep pairs that aligned well with each other close together to emphasize patterns in their alignments with the reference networks. Darker colors indicate better alignments.}  
    \end{center}
\end{figure}

\section{Discussion}

We have developed an algorithm that aligns networks using diffusion kernels to simulate temporal dynamics. This approach was shown to characterize nodes in a network in accordance with common measures of centrality, and to behave smoothly with respect to the number of changes in a network's edges allowing system trajectories to be captured. We successfully mapped a functional state space of four common network models and nine types of biological systems across a range of mechanisms and sizes, and were able to predict the underlying process in a single-model network of unknown origin. Unlike most statistical inference which aligns data from a single system to a proposed model, our approach need not even involve a model. Indeed, it is more akin to machine learning in that it looks for similarities in patterns of network dynamics across many pairwise comparisons offering a nonparametric means of tracking and classifying complex systems.

Why attempt to align networks by simulating dynamics on static models inferred from data that was never intended to capture dynamics so literally? Because network data with a temporal component is often challenging to collect so most network data are currently static. These data may only address the structure of a system's direct interactions, but these can in turn shed light on the more complicated effects of indirect interactions which have been shown to play a critical role in the assembly and control of biological systems \cite{matsui2000direct, wootton1994predicting, schuster2006network, kahramanoglou2011direct}. Our approach is an attempt to infer systemic functioning from the same common network data. 

The graph and subgraph isomorphism problems have been adopted by network scientists such that measures of alignment quality still emphasize topology over dynamics or functioning. From a practical perspective, of needing to assess and predict system trajectories, measures like ICS seem to hold less potential. If indirect effects matter as much as direct ones, then global network alignment ought to be reformulated away from data-centric topological models toward ones of system dynamics. Consider an invasive species like the zebra mussel. Physiologically, even genetically, zebra mussels vary little between the Eurasian bodies of water they are native to and the North American lakes and rivers they have invaded \cite{stepien2002genetic}. However from a systemic perspective they are different species. Similarly, two very distinct species can function almost identically in their respective communities. This idea was most famously formulated in the concept of a keystone species decades ago by Robert Paine \cite{paine1966food, paine1969note}. We would argue that just as there are functionally similar species, there are also functionally similar communities. Moreover, this is just as true in microbiology as ecology. It is encouraging to see that aligning networks by their dynamics captured some of their underlying functioning, even when the dynamics were simulated using static data.

There are however network aligners which were designed to generate heuristic solutions to the graph and subgraph isomorphisms problems. The two networks in Figure 1 illustrate why our approach should be used cautiously in this context. While diffusion from each node in Network 1 produced a unique pattern of entropy over time, only one node in Network 2 did. The trajectories of diffusion kernels emanating from nodes 2, 3, and 4 in the second network are identical making them all pairwise interchangeable in the final alignment. Our analyses used only a single best alignment, but considering the uniqueness of this alignment may offer additional insight into its inferential and predictive utility. Similarly, considering the probability of an alignment score given the two networks being aligned may offer additional insight. However this involves aligning permutations of the two networks to bootstrap an empirical distribution of alignment scores which is computationally expensive and requires justifying a particular permutation procedure.    

It is important to note a coupled benefit and drawback of the simulation-oriented approach developed here. Most other approaches to aligning networks assume static data, where the node and edge properties are fixed. This is logistically and analytically convenient. The approach developed here intentionally prioritizes the underlying conceptual approach over its computational complexity, which is rate limited by two steps. The Hungarian algorithm runs in cubic time with the number of nodes in a network \cite{edmonds1972theoretical}, though this can be substituted for any heuristic solution to the assignment problem \cite{clark2014comparison} which leaves matrix exponentiation as the rate-limiting algorithmic step. The current best upper bound on the complexity of matrix multiplication is $\mathcal{O}(\textrm{nodes}^{2.37})$ \cite{le2014powers}. This is only a modest improvement on the almost three decades old algorithm published by Coppersmith and Winograd \cite{coppersmith1990matrix} and therefore seems unlikely to improve dramatically in the future. In practice this limits our approach to networks of a few thousand nodes. However, the assumption that network data is static has already limited the analysis of noisier systems where interactions are continually changing \cite{berlow2004interaction}. Our approach can be used to align networks with explicit temporal components by having node and/or edge properties change mid-diffusion. The same approach can also be used to model communities more stochastically. The diffusion kernels used here are only the deterministic mean-field approximations of the true stochastic interactions. If those interactions have an asymmetric distribution of strengths, Jensen's inequality \cite{jensen1906fonctions} proves the mean rates of diffusion between nodes will give a biased view of the system's dynamics. These limitations make a diffusion-based alignment more tractable alongside the long-term goals of network science.

Our findings argue that as network science continues to grow as a field, encompassing an increasingly diverse set of biological systems as well as physical and social communities, there will be a growing need for unification. The practicality of this study's underlying goal to identify different types of systemic functioning using statistical network dynamics will depend heavily on the way network scientists incorporate biases into their data collection and analyses. There is no single correct way to compare networks so it is important to decide what kind of comparison will be most insightful relative to the questions being asked of the networks before comparing them. As an example, consider the goal of making size-invariant comparisons and what role a system's size might actually play in its functioning. For models like the one proposed by Erdos and Renyi \cite{erdos1959random} there is an assumed size-invariance because the density of edges is both random and size-invariant. However some systems are actually not size-invariant. Consider two systems formed through preferential attachment, one young and the other old. We would like to classify them similarly regardless of their age, but they function quite differently. Time will have turned the older system into a less equitable system that spreads information (or energy or any currency of interest) in a more predictable top-down manner. They are the result of the same process, but should they be classified together? Network scientists need to better understand how tools like an alignment algorithm are actually used, and agree more on the ways they envision them shedding light on systems of all kinds.

\section*{Acknowledgments}
The authors would like to thank the Interdisciplinary Quantitative Biology program at CU Boulder for instigating this collaboration, Dan F. Doak for imbuing it with clarity and utility, Andrew D. Orr for help turning ideas into code, and Joseph G. Azofeifa for years of insightful questioning and encouragement. Funding was provided by the following National Science Foundation grants: IGERT 1144807, GK-12 0841423, and DGE-1144083. The authors declare no competing interests.

\bibliographystyle{plain}
\bibliography{NetCom}

\clearpage

\beginsupplement

\section*{Supplementary Information}

\begin{table}[H]
    \begin{center}
      \includegraphics[scale=0.5]{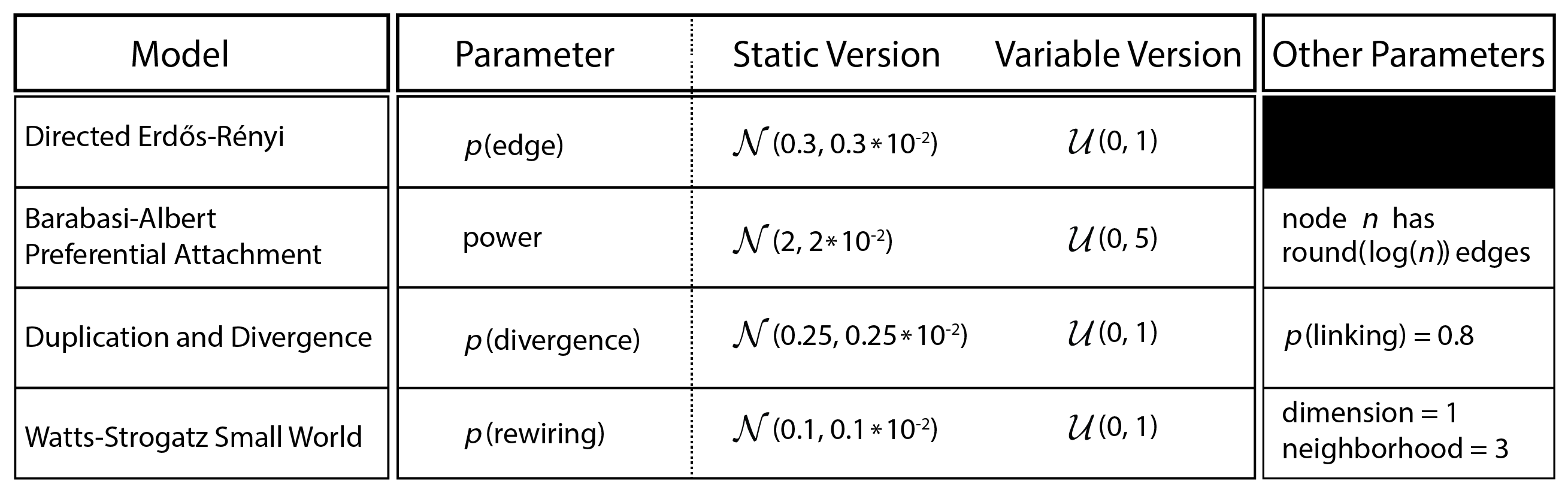} 
      \caption{Model specifications for synthetic networks.}
    \end{center}
\end{table}

\begin{table}[H]
    \begin{center}
      \includegraphics[scale=0.8]{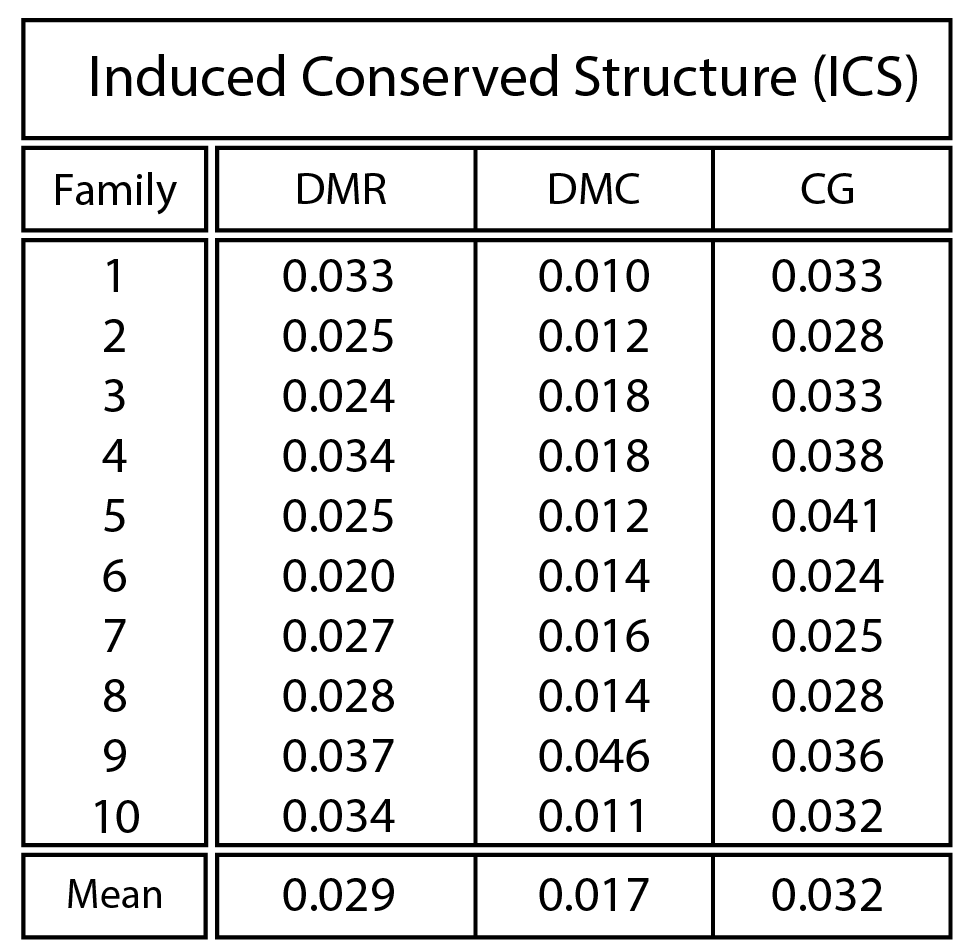} 
      \caption{Induced Conserved Structure (ICS) scores for the alignments produced by the method developed here when applied to the same 30 pairs of networks recently compared by Clark and Kalita \cite{clark2014comparison}. The three network families were Duplication with Random Mutation (DMR), Duplication-Mutation-Complementation (DMC), and Crystal Growth (CG). Each pair of networks was comprised of one 3000 node network and one 4000 node network.}
    \end{center}
\end{table}

\begin{figure}[H]
      \includegraphics[max width=\linewidth]{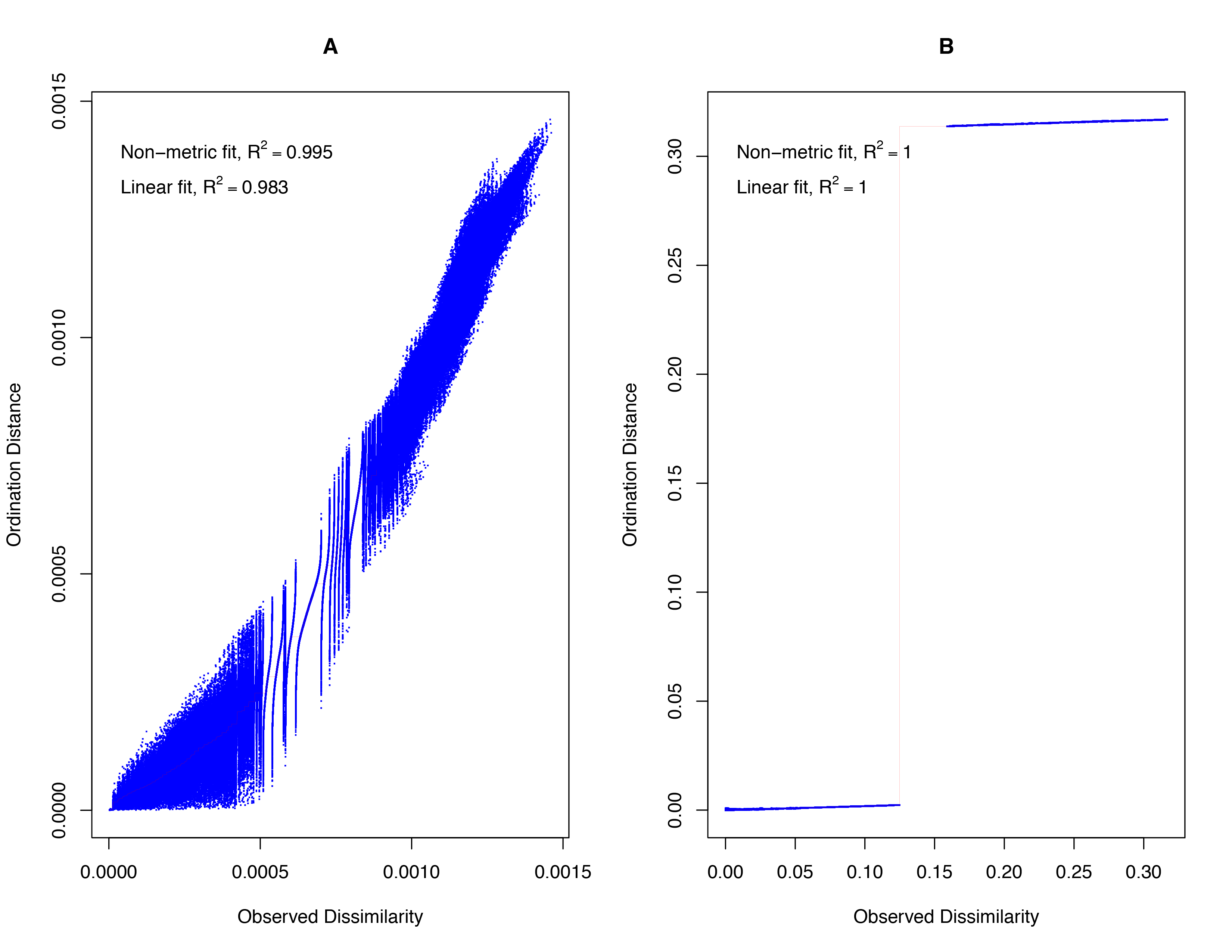} 
      \caption{Shepard plots for the non-metric multidimensional scaling of network dynamics in Figure 5. (A) Stress = 0.07. (B) Stress = 0.0002.}
\end{figure}

\afterpage{
\begin{longtable}[H]{| p{.2\textwidth} | L{.2\textwidth}   L{.3\textwidth}   L{.3\textwidth} |} 
            \hline
                \multicolumn{1}{|c|}{ID} & \multicolumn{1}{c}{Classification} & \multicolumn{1}{c}{Source} & \multicolumn{1}{c|}{Details} \\ 
            \hline
            \hline
                \multicolumn{1}{|c|}{ 1 - 30 } & Erdos-Renyi & R igraph package \cite{iGraph} & 10 each with 10, 100, and 1000 nodes from the Static Version in Table S1\\ 
            \hline    
                \multicolumn{1}{|c|}{ 31 - 60 } & Preferential Attachment & R igraph package \cite{iGraph} & 10 each with 10, 100, and 1000 nodes from the Static Version in Table S1\\ 
            \hline
                \multicolumn{1}{|c|}{ 61 - 90 } & Duplication and Divergence & \cite{ispolatov2005duplication, vazquez2002modeling, gibson2011improving} & 10 each with 10, 100, and 1000 nodes from the Static Version in Table S1\\
            \hline
                \multicolumn{1}{|c|}{ 91 - 120 } & Small World & R igraph package \cite{iGraph} & 10 each with 10, 100, and 1000 nodes from the Static Version in Table S1\\
            \hline
                \multicolumn{1}{|c|}{ 121 - 218 } & Gene Regulatory Networks & ENCODE Consortium \cite{encode2012integrated} and Dr. Yijun Ruan & 100 randomly selected connected components from the following Accession Numbers: ENCFF001THV, ENCFF001THX, ENCFF001THY, ENCFF001THZ, ENCFF001TIA, ENCFF001TIB, ENCFF001TID, ENCFF001TIE, ENCFF001TIF, ENCFF001TIJ, ENCFF001THU, ENCFF001THT, ENCFF001TIG, ENCFF001THW, ENCFF001TIC  \\
            \hline
                \multicolumn{1}{|c|}{ 219 - 276 } & Trophic Ecosystems & R enaR package \cite{borrett2014enar} & \\
            \hline
                \multicolumn{1}{|c|}{ 277 - 318 } & Biogeochemical Ecosystems & R enaR package \cite{borrett2014enar} & \\
            \hline
                \multicolumn{1}{|c|}{ 319 } & Macaque Brain & R igraphdata package \cite{negyessy2006prediction} & Visuotactile regions\\ 
            \hline
                \multicolumn{1}{|c|}{ 320 } & \textit{C. elegans} Metabolism & Network Repository \cite{nr, duch2005community} &  \\
            \hline
                \multicolumn{1}{|c|}{ 321 } & Human Diseasome & Network Repository \cite{nr, goh2007human} & \\
            \hline
                \multicolumn{1}{|c|}{ 322 } & Yeast Protein-Protein Interactions & Network Repository \cite{nr, jeong2001lethality} & \\
            \hline
                \multicolumn{1}{|c|}{ 323 } & Protein-Protein Interactions & R COSINE package\cite{COSINE} & \\
            \hline
                \multicolumn{1}{|c|}{ 324 - 327 } & Protein-Gene Interactions & R yeastExpData package\cite{yeastExpData} & \\
            \hline    
                \multicolumn{1}{|c|}{ 328 - 427 } & Enzymatic Pathways & Network Repository Cheminformatics \cite{nr} & g2, g12, g30, g35, g39, g54, g57, g59, g62, g64, g70, g72, g97, g101, g115, g117, g121, g122, g129, g132, g137, g141, g142, g144, g149, g158, g162, g167, g170, g171, g175, g189, g192, g201, g203, g204, g206, g208, g217, g230, g245, g246, g247, g259, g263, g264, g275, g298, g299, g302, g304, g310, g314, g321, g332, g337, g345, g354, g362, g363, g365, g375, g377, g378, g387, g392, g397, g401, g410, g412, g415, g421, g423, g438, g439, g444, g458, g459, g464, g474, g481, g509, g516, g520, g532, g533, g535, g537, g543, g544, g558, g560, g562, g563, g574, g578, g588, g592, g596, g598 \\
            \hline
    \caption{Sources for all 427 networks in Figure 6. All networks were freely available as of the publication date.}
\end{longtable}
}

\begin{figure}
    \begin{center}
      \includegraphics[max width=\linewidth]{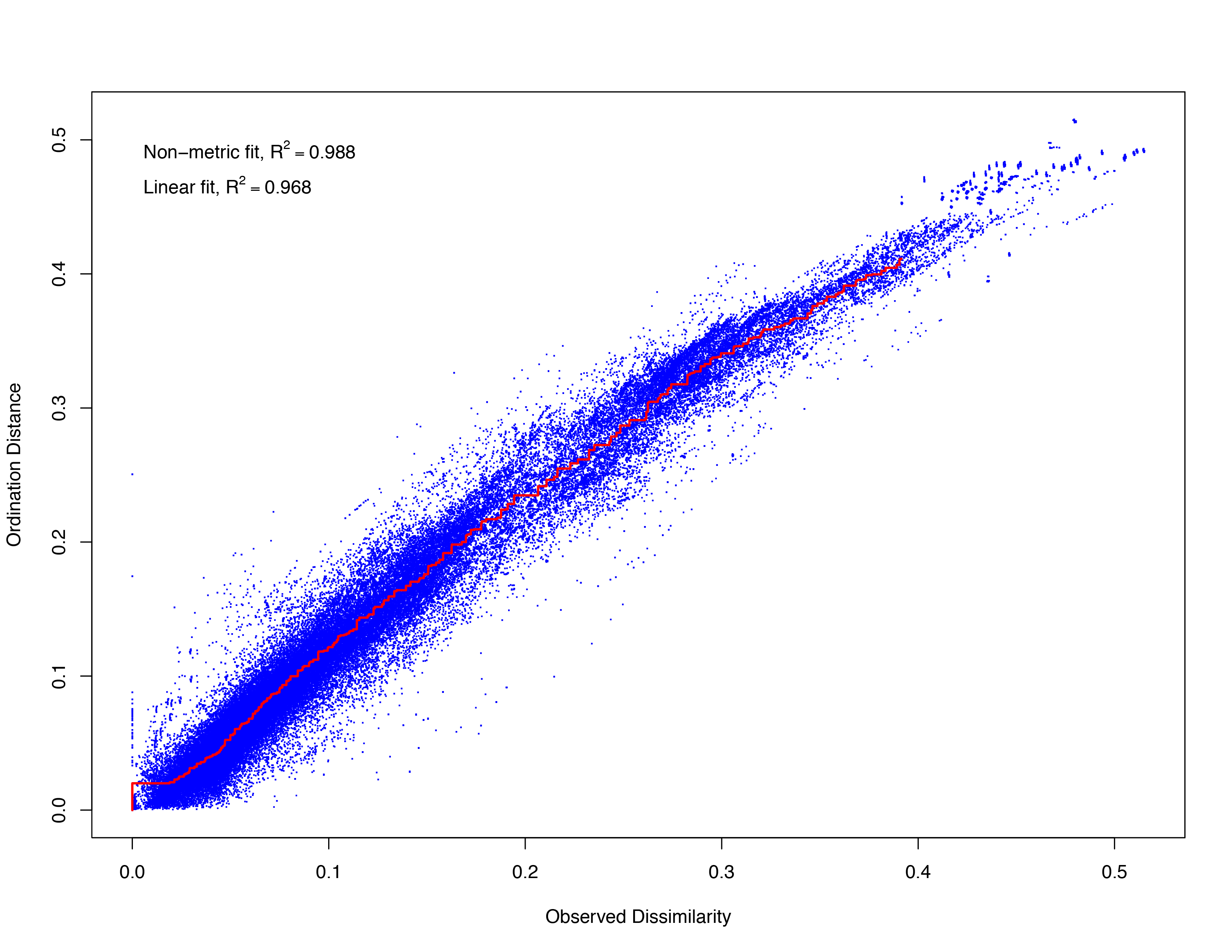} 
      \caption{Shepard plot for the non-metetric multidimensional scaling of pairwise alignments between the 307 empirical networks and 120 reference networks in Figure 6. Stress = 0.11.}
    \end{center}
\end{figure}

\begin{figure}
    \begin{center}
      \includegraphics[max width=\linewidth]{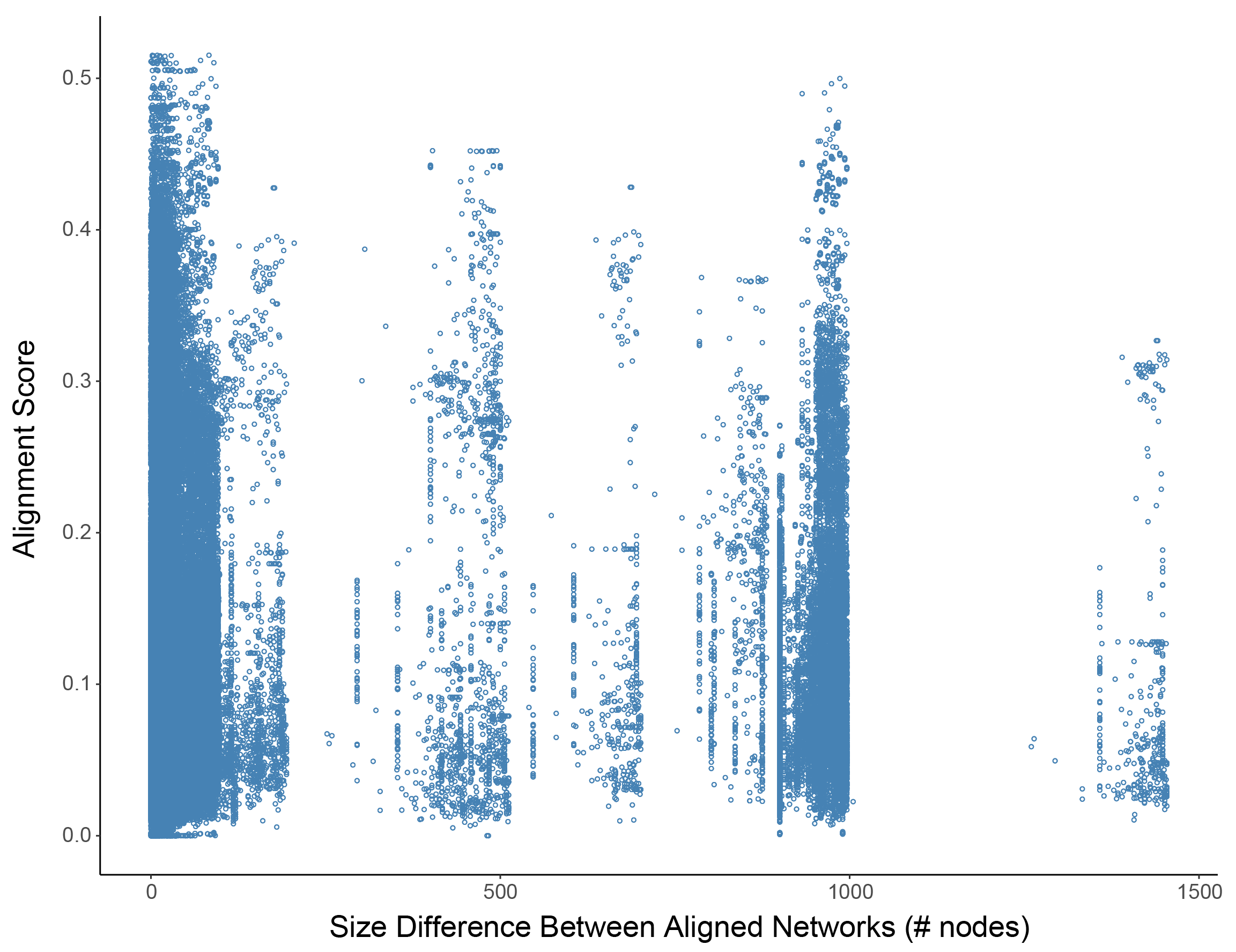} 
      \caption{Relationship between the difference in network size (nodes) and the quality of their alignment, indicating the alignments were size-invariant.}
    \end{center}
\end{figure}

\begin{figure}
    \begin{center}
      \includegraphics[max width=\linewidth]{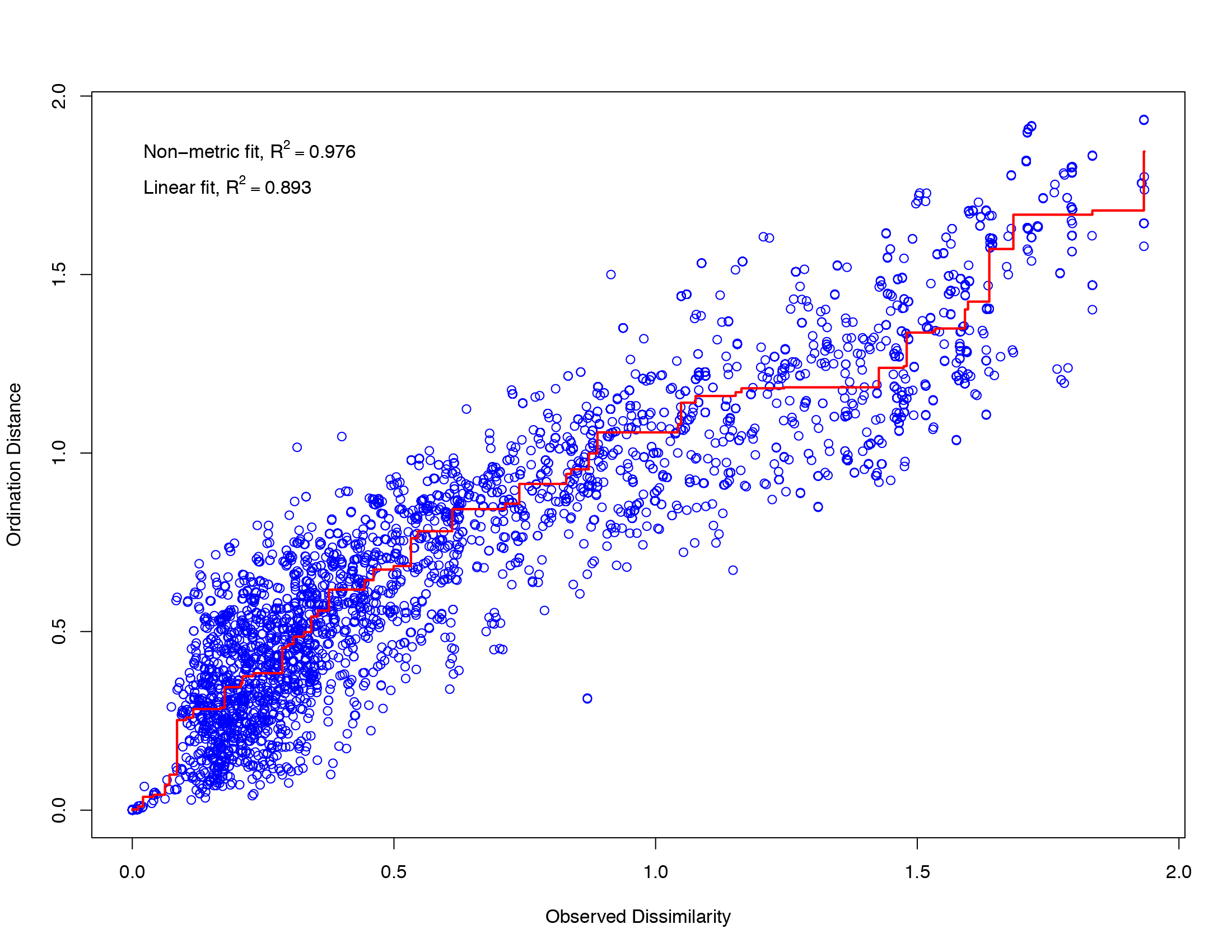} 
      \caption{Shepard plot for the non-metric multidimensional scaling of pairwise alignments between the 100 empirical ecological networks in Figure 5. Stress = 0.16.}
    \end{center}
\end{figure}

\end{document}